\documentclass[twocolumn,prb,groupedaddress,preprintnumbers,amsmath,amssymb, floatfix]{revtex4}

\usepackage{amsmath}
\usepackage{bm}
\usepackage{graphicx}
\usepackage{dcolumn}
\usepackage{bm}
\begin{document}

\title{Structure and interaction of flexible dendrimers in concentrated solution} 

\author{S. Rosenfeldt, M.~Ballauff}
\affiliation{Physical Chemistry I, University of Bayreuth, 95440 Bayreuth, Germany}
\author{P.~Lindner}
\affiliation{Institut Laue-Langevin, B.P.\ 156X, 38042 Grenoble Cedex, France}
\author{L.~Harnau$^*$}
\affiliation{Max-Planck-Institut f\"ur Metallforschung, Heisenbergstr. 3, D-70569 Stuttgart, 
Germany, \\and Institut f\"ur Theoretische und Angewandte Physik, Universit\"at Stuttgart, 
Pfaffenwaldring 57, D-70569 Stuttgart, Germany}
\email[]{E-mail: harnau@fluids.mpi-stuttgart.mpg.de} 
\date{\today}

\begin{abstract} 
We study the influence of mutual interaction on the conformation of flexible 
poly(propyleneamine) dendrimers of fourth generation in concentrated solution.
Mixtures of dendrimers with protonated and deuterated end groups are investigated by 
small-angle neutron scattering up to volume fractions of 0.23. This value is in the 
range of the overlap concentration of the dendrimers. The contrast between the solute 
and the solvent was varied by using mixtures of protonated and deuterated solvents.
This allows us to investigate the partial structure factors of the deuterated dendrimers
in detail. An analysis of the measured scattering intensities reveals that the shape
of the flexible dendrimers is practically independent of the concentration in contrast to the pronounced conformational changes of flexible linear polymers.
\end{abstract}
\maketitle

\section{Introduction}
\begin{figure}[t]
\includegraphics[width=7.8cm,clip]{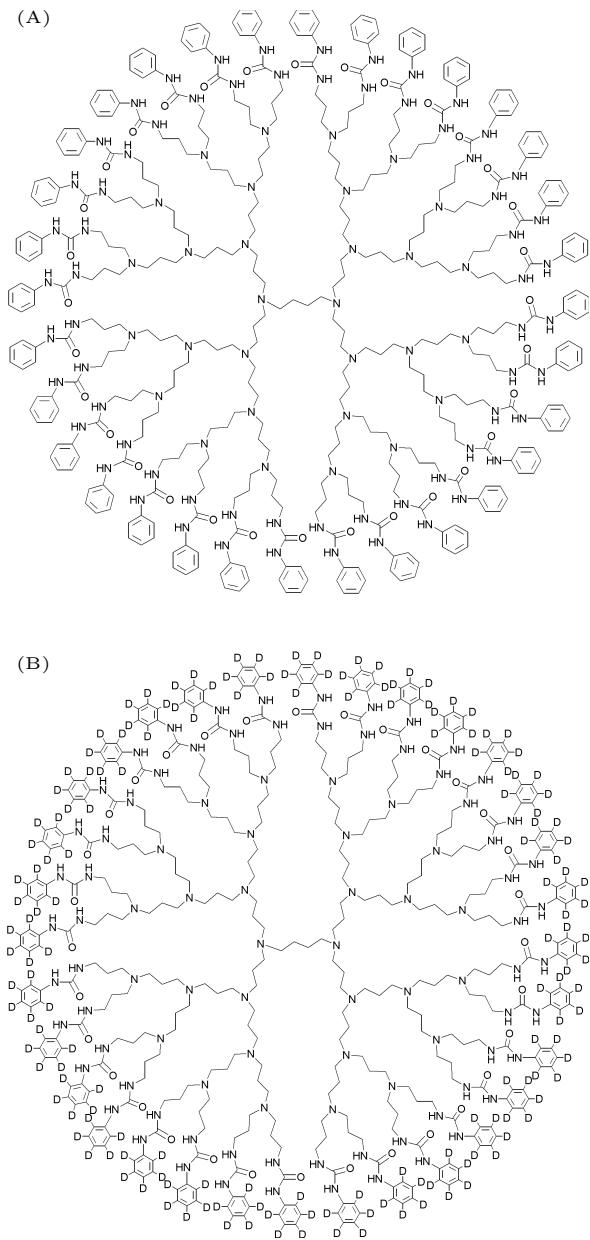}
\caption{Chemical structure of the functionalized poly(propyleneamine) dendrimer
of fourth generation \mbox{G4-H} having protonated end groups in (A) and the 
same dendrimer but with deuterated end groups \mbox{G4-D} in (B).}
\label{fig1}
\end{figure}

Dendrimers are macromolecular structures that exhibit a defined tree-like architecture. 
\cite{fich:99,ball:04,meij:03,toma:05} Figure \ref{fig1} displays a flexible
poly(propyleneamine) dendrimer of fourth generation. The subsequent units are emanating
from a focal unit and a monodisperse macromolecular structure results. In this way,
dendrimers combine properties of polymers and colloids \cite{ball:04} and may find
applications such as contrast agents in magnetic resonance imaging, \cite{regi:08,krau:00}
light-harvesting systems, \cite{kwon:07} and drug-delivery systems. \cite{abhi:08,shab:04}

Up to now, the spatial structure of dendrimers is fully understood in the limit of infinite 
dilution. Figure \ref{fig1} immediately demonstrates that flexible dendrimers possess a 
large number of conformational degrees of freedom which follow from rotations about 
various chemical bonds. Hence, the terminal groups can fold back. As a result of these
conformational degrees of freedom, flexible dendrimers adopt a dense core structure (see
the discussion in Refs.~\cite{ball:04,rose:02a,rose:02b,ding:03}).
On the other hand, no backfolding of the terminal groups can occur in the case of rigid
dendrimers such as polyphenylene dendrimers \cite{rose:04,rose:05} or stilbenoid 
dendrimers. \cite{rose:06} 

Much less is known about the structure and interaction of flexible dendrimers in
concentrated solutions.  Compared to the huge number of papers on dendrimers in general, 
only a small number of theoretical  \cite{liko:01,liko:02,goet:03,harr:03,goet:04,goet:05,goet:06,liko:07,tera:07,tera:07a,harn:07,mladek:08}
and experimental studies \cite{topp:99,bodn:00,uppu:98,sagi:02,rose:02b} focus on 
properties of dendrimers at higher concentrations. Thus, Topp et al. \cite{topp:99} 
studied solutions of poly(propyleneimine) dendrimers (mass fractions between 0.01 and 0.80) 
by small-angle scattering techniques. They concluded that the size of the dendrimers 
decreases upon increasing the dendrimer concentration. Bodnar et al. \cite{bodn:00}  
conducted small-angle neutron scattering and rheological measurements at higher concentrations
of poly(propyleneimine) dendrimers. These authors suggest dendrimer clustering and 
interpenetration with increasing concentration. Sagidullin et al. \cite{sagi:02} studied 
the self-diffusion coefficient of flexible dendrimers up to high volume fractions and 
obtained the respective scaling exponents.

The first theoretical treatment of the interaction of flexible dendrimers is due to Likos and 
coworkers. \cite{liko:01} Starting from the Gaussian density distribution of a flexible 
dendrimer of fourth generation, they demonstrated that the interaction potential $U(r)$ can 
be directly derived from this density distribution. For not too concentrated solution $U(r)$ 
has been modeled by a Gaussian in good agreement with experimental data. \cite{liko:02} A 
significant difference between the measured and the calculated structure factor was only 
observed at the highest concentration. Goetze et al. \cite{goet:05} employed monomer resolved 
computer simulations to examine the significance of many-body effects in concentrated dendrimer 
solutions. They reported that the effects of many-body forces are small in general and become 
weaker as the dendrimer flexibility increases. More recently, this problem was re-considered 
by Terao who performed coarse-grained molecular dynamics simulations of charged dendrimers 
in aqueous solution in order to clarify the influence of many-body interactions in concentrated 
solution. \cite{tera:07,tera:07a} Mladek et al. \cite{mladek:08} showed that amphiphilic 
dendrimers form clusters of overlapping particles in the fluid, which upon further compression 
crystallize into cubic lattices with density-independent lattice constants.

Summarizing the work done so far only Ref.~\cite{liko:02} has given a quantitative comparison
between theory and experiment. It seems fair to say that a comprehensive study of the
interaction of flexible dendrimers in concentrated solution is still missing.
Moreover, we lack further insight into possible distortions of the structure of flexible
dendrimers due to mutual interaction. Many soft materials such as flexible polymers
(see, e.g., Refs.~\cite{daou:75,yeth:91}), bottle-brush polymers (see, e.g.,
Refs.~\cite{boli:07,boli:08}), semiflexible polyelectrolytes (see, e.g.,
Refs.~\cite{stev:93,yeth:97}), polyelectrolyte brushes (see, e.g., Refs.~\cite{xu:09,yan:09}),
and microgels (see, e.g., Refs.~\cite{ball:07,boli:09}) can be
induced to change their shape due to various interactions. Up to now, a similar analysis of possible
shape distortions of flexible dendrimers at high concentrations is still missing. 

Here we present a study of the interaction of flexible dendrimers of fourth generation in 
concentrated solution. As the method, we chose small-angle neutron scattering (SANS). 
\cite{higg:94} Small-angle scattering is the method of choice because it allows us to study 
the spatial correlations that arise from the excluded volume interactions between the monomers 
of different dendrimers in the solution. Partially deuterated dendrimers are mixed with the 
same but protonated dendrimers in dimethylacetamide (DMA) which is a good solvent for this 
system.  As in our previous study of a one-component system \cite{rose:02a} contrast variation 
between the solute dendrimer and the solvent is used by use of mixtures of protonated and the deuterated (DMA). In this way the partial scattering intensities of the deuterated  species becomes available. This allows us to investigate
both the {\it structure} and the {\it interaction} of dendrimers in concentrated solution. In particular, such a study can clearly reveal whether the conformation of dendrimers will be distorted by mutual interaction in concentrated solution.

\begin{table}[t]
\begin{tabular}{|c|c|c|c|c|} \hline
sample 	& $\phi $ & $\phi_H$  & $\phi_D$ &  contrast \\ \hline \hline 
$S_1$	& 0.15	& 0.12	& 0.03		 &  G4-H	  \\ \hline
$S_2$	& 0.15	& 0.15	& 0		 &		  \\ \hline
$S_3$	& 0.23	& 0.03	& 0.20		 &  G4-D	  \\ \hline	
$S_4$	& 0.23  & 0.23	& 0		 &	     \\ \hline
\end{tabular}
\caption{Overview of the investigated samples. The volume fractions of the protonated 
and partially deuterated dendrimers are denoted as $\phi_H$ and $\phi_D$, respectively.
$\phi=\phi_H+\phi_D$ is the total volume fraction of the dendrimers. 
The scattering length density of the solvent (contrast) has been chosen such 
that samples $S_1$  and $S_3$ are investigated at the match point of the G4-H and 
G4-D dendrimers, respectively.}
\label{table:1}	
\end{table}
\section{Experimental Section}
The G4-H dendrimer was synthesized according to Ref. \cite{step:99}. The specific
volume of the G4-H dendrimer was determined to be $\bar{\nu}=0.84 \pm 0.01 \,
{\rm cm^3/g}$ using a DMA-60-densitometer (Paar, Graz, Austria). The partially
deuterated dendrimer G4-D was synthesized using fully deuterated phenylisocyanate.
The specific volume of the G4-D dendrimer has the value $\bar{\nu}=0.82 \pm 0.01 \,
{\rm cm^3/g}$. Protonated DMA (Fluka, analytical grade) and deuterated DMA (DMA-$d_9$,
degree of deuteration 99\%, Deutero GmbH) were used as received.

We investigated various mixtures of flexible dendrimers of fourth generation
with protonated and deuterated end groups by SANS. An overview of the samples 
is given in Table \ref{table:1}, in which we also indicate if the measurements 
have been performed at the match point of a dendrimer.

The SANS measurements were done using the instrument D11 at the Institut 
Laue-Langevin. Software provided at the instrument were used to obtain the 
radial averaged intensities in absolute scale. \cite{lind:92,lieu:07} Further 
data treatment was carried out according to literature procedures. 
\cite{rose:02a,rose:02b,rose:06} For all data shown here, the rates of 
incoherent scattering caused by the protons were determined at high 
scattering vector, set as a constant and subtracted from the bare data.

\section{Theory}
\subsection{Scattering intensity and Ornstein-Zernike equations}

SANS determines the scattering intensity $I({\bf q})$ as a function of the 
scattering vector ${\bf q}$ and the concentration of the dissolved particles. 
In addition to the coherent scattering $I_{coh}({\bf q})$, there is always an 
incoherent contribution $I_{incoh}$ that is due to the protons present in the 
particles under consideration. The scattering intensity can be written as 
\begin{eqnarray} \label{eq1}
I({\bf q})&=&I_{coh}({\bf q})+I_{incoh}\,.
\end{eqnarray}
Note that in the notation the dependence on the concentration is suppressed 
The ${\bf q}$-independent incoherent contribution $I_{incoh}$ of individual 
particles must be subtracted carefully from experimental data in order to obtain 
meaningful results on the structure and interaction of the dissolved particles. 
\cite{rose:02b} The systems under consideration are solutions. In view of the 
mesoscopic scale of the particles, the solvent will be modeled as structureless 
continuum providing a homogeneous scattering length density.

In general we consider a multicomponent system involving $\nu$ species of particles
with particle numbers $N_H$ in the volume $V$. Each particle of a species $H$
($1 \le H \le \nu$) carries $n_H$ scattering units. The coherent contribution to
the scattering intensity in the $\nu$-component system is given by
\begin{eqnarray}  \label{eq2}
I_{coh}({\bf q})&=&\sum\limits_{H=1}^\nu \sum\limits_{D=1}^\nu I_{HD}({\bf q})\,,
\end{eqnarray}
with the partial scattering intensities 
\begin{eqnarray}   
I_{HD}({\bf q})&=&\frac{1}{V}\left\langle
\sum\limits_{i=1}^{n_H}\sum\limits_{j=1}^{n_D}
\sum\limits_{\alpha=1}^{N_H}\sum\limits_{\gamma=1}^{N_D}
b_{iH}^{(\alpha)} b_{jD}^{(\gamma)}
e^{i{\bf q}\cdot \left({\bf r}_{iH}^{(\alpha)}-{\bf r}_{jD}^{(\gamma)}\right)}
\right\rangle\,.\nonumber
\\&&            \label{eq3}
\end{eqnarray}
Here  ${\bf r}_{iH}^{(\alpha)}$  is the position vector of the $i$-th scattering 
unit ($1 \le i \le n_H$) of the $\alpha$-th particle ($1 \le \alpha \le N_H$)
of species $H$. The difference of the scattering length of this scattering unit 
and the average scattering length of the solvent is denoted as $b_{iH}^{(\alpha)}$,
and $\langle \,\,\,\, \rangle$ is an ensemble average.

It proves convenient to decompose the partial scattering intensities according to 
\begin{eqnarray}   \label{eq4}
I_{HD}({\bf q})&=&{\tilde \rho}_H {\tilde \omega}_H({\bf q})\delta_{HD}
+{\tilde \rho}_H {\tilde \rho}_D {\tilde h}_{HD}({\bf q})\,,
\end{eqnarray}
where
\begin{eqnarray}   \label{eq5}
{\tilde h}_{HD}({\bf q})&=&\frac{V}{N_H N_D n_H n_D}\nonumber
\\\times&&\!\!\!\!\!\!\!\!\left\langle
\sum\limits_{i=1}^{n_H}\sum\limits_{j=1}^{n_D}
\sum\limits_{\alpha=1}^{N_H}\sum\limits_{\stackrel{\gamma=1}{\gamma\neq \alpha}}^{N_D}
b_{iH}^{(\alpha)} b_{jD}^{(\gamma)}
e^{i{\bf q}\cdot \left({\bf r}_{iH}^{(\alpha)}-{\bf r}_{jD}^{(\gamma)}\right)}
\right\rangle
\end{eqnarray}
is a particle-averaged total correlation function for pairs of particles of species 
$H$ and $D$. The scattering unit number density of particles of species $H$ is designated
by ${\tilde \rho}_H=N_H n_H /V$. The particle-averaged intramolecular correlation function
\begin{eqnarray}   \label{eq6}
{\tilde \omega}_H({\bf q})&=&\frac{1}{N_H n_H}\left\langle
\sum\limits_{i=1}^{n_H} \sum\limits_{j=1}^{n_H} \sum\limits_{\alpha=1}^{N_H}
b_{iH}^{(\alpha)} b_{jH}^{(\alpha)}
e^{i{\bf q}\cdot \left({\bf r}_{iH}^{(\alpha)}-{\bf r}_{jH}^{(\alpha)}\right)}
\right\rangle\nonumber
\\&&
\end{eqnarray}
characterizes the scattering length distribution, and hence also the geometric shape 
of particles of species $H$. While the particle-averaged intramolecular correlations 
functions account for the interference of radiation scattered from different parts of 
the same particle in a scattering experiment, the local order 
in the fluid is characterized by the total correlation functions.
In the case of chemically homogeneous particles characterized by $b_H=b_{iH}^{(\alpha)}$, 
the partial scattering intensities defined in Eq.~(\ref{eq4}) can be written as 
\begin{eqnarray}  
I_{HD}({\bf q})&=&{\tilde \rho}_H (b_H)^2\omega_H({\bf q})\delta_{HD}
+{\tilde \rho}_H {\tilde \rho}_D   b_H b_D h_{HD}({\bf q})\nonumber
\\&& \label{eq7}
\end{eqnarray}
with $(b_H)^2\omega_H({\bf q})={\tilde \omega}_H({\bf q})$ and 
$b_H b_D h_{HD}({\bf q})={\tilde h}_{HD}({\bf q})$.
The total correlation functions $h_{HD}({\bf q})$ are related to a set of 
direct correlation functions $c_{HD}({\bf q})$ by generalized Ornstein-Zernike 
equations of the polymer reference interaction site model (PRISM), which in 
Fourier space read \cite{schw:97,harn:08,yeth:09}
\begin{eqnarray}  
h_{HD}({\bf q})&\!\!=\!\!&\sum\limits_{A=1}^\nu \omega_H({\bf q}) c_{HA}({\bf q})
\left(\omega_A({\bf q})\delta_{AD}+{\tilde \rho}_A h_{AD}({\bf q})\right).
\nonumber
\\&& \label{eq8}
\end{eqnarray}
This set of generalized Ornstein-Zernike equations must be supplemented by a 
set of closure equations. The PRISM integral equation theory has been 
successfully applied to various experimental systems such as monodisperse 
($\nu = 1$) polymers \cite{yeth:96,harn:01a,boli:07} and rigid dendrimers, 
\cite{rose:06,harn:07} binary mixtures ($\nu = 2$) of charged colloids,  
\cite{shew:98,harn:02} three-component mixtures ($\nu = 3$) of charged colloids 
and salt ions, \cite{harn:00,harn:01} as well as polydisperse ($\nu \gg 1$) 
nanoparticles  \cite{webe:07} and polyelectrolyte brushes. \cite{henz:08}

\subsection{Decoupling approximation}
A common decoupling approximation is to assume that the
conformations of two particles are independent of each other and of the 
two mutual positions of their center-of-mass ($cm$), such that the statistical 
average in Eq.~(\ref{eq5}) factorizes. In this case the coherent contribution 
to the scattering intensity for a binary mixture ($\nu = 2$) is given by 
\begin{eqnarray}  \label{eq9}
I_{coh}({\bf q})&=&\rho_H I^{(0)}_H({\bf q}) 
\left(1+\rho_H h^{(cm)}_{HH}({\bf q})\right)\nonumber
\\&+&\rho_D I^{(0)}_D({\bf q}) 
\left(1+\rho_D h^{(cm)}_{DD}({\bf q})\right)\nonumber
\\&+&2\rho_H\rho_D\sqrt{I^{(0)}_H({\bf q})I^{(0)}_D({\bf q})}
h^{(cm)}_{HD}({\bf q})\,,
\end{eqnarray}
where $\rho_H=N_H/V$ is the number density of particles of species $H$.
The scattering intensity of a single particle of species $H$ reads 
\begin{eqnarray}  
I^{(0)}_H({\bf q})&=&\frac{1}{N_H}\sum\limits_{\alpha=1}^{N_H}
\left\langle\sum\limits_{i=1}^{n_H} b_{iH}^{(\alpha)} 
e^{i{\bf q}\cdot {\bf l}_{iH}^{(\alpha)}}\right\rangle
\left\langle\sum\limits_{j=1}^{n_H} b_{jH}^{(\alpha)} 
e^{i{\bf q}\cdot {\bf l}_{jH}^{(\alpha)}}\right\rangle\,,
\nonumber
\\&& \label{eq10}
\end{eqnarray}
where the position vector of the $i$-th scattering unit on the $\alpha$-th 
particle of species $H$ is written as ${\bf r}_{iH}^{(\alpha)}=
{\bf R}_{H}^{(\alpha)}+{\bf l}_{iH}^{(\alpha)}$. Here 
${\bf R}_{H}^{(\alpha)}$ is the $cm$ position vector of the $\alpha$-th 
particle of species $H$. Spatial pair correlations are 
characterized by a set of $cm-cm$ total correlation functions 
\begin{eqnarray}  \label{eq11}
h^{(cm)}_{HD}({\bf q})&=&\frac{V}{N_HN_D}
\sum\limits_{\alpha=1}^{N_H}
\sum\limits_{\gamma=1}^{N_D}
\left\langle
e^{i{\bf q}\cdot \left({\bf R}_{H}^{(\alpha)}-{\bf R}_{D}^{(\gamma)}\right)}
\right\rangle.
\end{eqnarray}
These total correlation functions are related to $cm-cm$ direct correlation 
functions by the Ornstein-Zernike relations \cite{goet:06}
\begin{eqnarray}   \label{eq12}
h_{HD}^{(cm)}({\bf q})&\!\!\!=\!\!\!\!\!
\sum\limits_{A\in \{H,D\}}  c_{HA}^{(cm)}({\bf q})
\left(\delta_{AD}+\rho_A h_{AD}^{(cm)}({\bf q})\right).
\end{eqnarray}
The $cm-cm$ correlation functions can be expressed in terms of correlation 
functions between interaction sites by considering the $cm$ as an auxiliary 
site within the PRISM. \cite{krak:02,harn:02} Such a relationship provides 
a direct link between the correlation functions calculated from the PRISM 
theory and the $cm-cm$ correlation functions which are the basic input into 
the coarse-graining scheme. In particular the validity of the decoupling 
approximation can be investigated for particles of arbitrary shape by 
comparing the calculated scattering intensities with the results of the 
PRISM. \cite{krak:02} The decoupling approximation together with the 
Ornstein-Zernike relation [Eq. (\ref{eq12})] has been successfully applied to
monodisperse ($\rho_D=0$) poly\-(propylene\-amine) dendrimers of the fourth 
generation \cite{liko:02} although Eqs.~(\ref{eq9}) - (\ref{eq11}) are based 
on uncontrolled factorization approximations. A computer simulation study 
\cite{goet:05} has revealed a breakdown of this factorization approximation 
at high dendrimer particle number densities similar to earlier findings for 
flexible polymers based on a comparison with the results for the PRISM. 
\cite{krak:02} Starting around the overlap concentration, the following effects 
are expected: (a) The presence of many dendrimers surrounding a given one leads 
to a deformation of the dendrimer itself. With increasing concentration the 
spatial structure of a single dendrimer is expected to derivate more and more 
from the one given at infinite dilution. (b) Concerning the intermolecular 
correlations between the dendrimers many-body-correlations become more likely 
upon increasing the concentration.

\section{Results and discussion}
The systems under investigation are binary mixtures ($\nu=2$) of functionalized 
poly(propyleneamine) dendrimers of the fourth generation with protonated end 
groups (G4-H) denoted as dendrimers of species $H$ [see Fig.~\ref{fig1} (A)]. The 
same dendrimers with deuterated end groups (G4-D) are denoted as dendrimers of 
species $D$ [see Fig.~\ref{fig1} (B)]. All phenyl end groups of the G4-D dendrimers
have been fully deuterated to enhance their scattering length as compared to the 
dendritic scaffold. \cite{rose:02a} Note that the overall structure of 
both dendrimers is identical in the limit $\phi \to 0$  and the solvent DMA is a 
good solvent for both systems. \cite{rose:02a} Therefore, the intramolecular 
excluded volume interaction between the monomers significantly influences the 
shape of the dendrimers.

\subsection{Scattering intensity of single dendrimers}
As a prerequisite for studies on the coherent contribution $I_{coh}({\bf q})$ to the 
scattering intensity at various concentrations, we discuss first the scattering 
intensities  $I^{(0)}_H({\bf q})$  and $I^{(0)}_D({\bf q})$  [see Eqs.~(\ref{eq9}) 
and (\ref{eq10})] which have been determined earlier. \cite{rose:02a,rose:02b,ding:03} 
It has been shown that these scattering intensities can be split into three parts 
according to 
\begin{eqnarray}   \label{eq13}
I^{(0)}_A(q)&=&{\tilde b}^2_A I_A^{(S)}(q)+
2{\tilde b}_A I_A^{(SI)}(q)+I_A^{(I)}(q)\,,\nonumber
\\&&A\in\{H, D\}\,,
\end{eqnarray}
where $q=|{\bf q}|$ and ${\tilde b}_A$ is the so-called contrast, i.e., the difference 
of the average scattering length density of dendrimers of species $A$ and the scattering 
length density of the solvent. In the case of SANS, ${\tilde b}_A$ is a
parameter that can be varied by mixing protonated and deuterated solvent 
(contrast variation). \cite{rose:02a} The term $I_A^{(S)}(q)$ is the Fourier-transform 
of a shape function $T_A(r)$ that describes the statistical average over all possible 
conformations of the dendrimers. This quantity has been compared directly to results 
of computer simulation studies. \cite{goet:03,harr:03} The term $I_A^{(I)}(q)$ 
is related to the scattering length inhomogeneity of the dendrimers, while the term
$I_A^{(SI)}(q)$  presents the cross term between the former contributions. \cite{rose:02a} 
For both types of dendrimers (G4-H and G4-D) the shape term $I_A^{(S)}(q)$ can be described by 
the Gaussian function

\begin{figure}[t]
\includegraphics[width=8.5cm,clip]{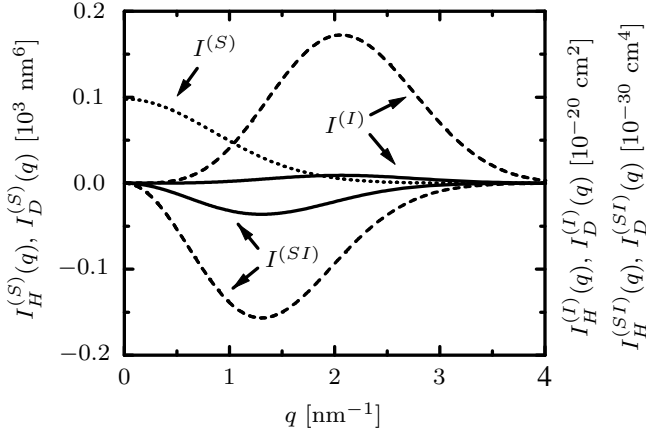}
\caption{The partial scattering intensities $I_H^{(I)}(q)$ [$I_H^{(SI)}(q)$]
of the G4-H dendrimer (upper [lower] solid line) and $I_D^{(I)}(q)$ [$I_D^{(SI)}(q)$]
of the G4-D dendrimer (upper [lower] dashed line)
together with $I_H^{(S)}(q) = I_D^{(S)}(q)$ (dotted line) as obtained from modeling 
scattering data from dilute solution \cite{rose:02a,ding:03} according to Eq.~(\ref{eq13}). 
An interpretation of these data leads to the following conclusions: The overall structure 
of both dendrimers is identical (dotted line), an appreciable number of end groups 
must fold back into the interior of the dendrimers (dashed lines), and the 
scattering length distribution is rather homogeneous within in the G4-H dendrimer
(solid lines).}
\label{fig2}
\end{figure}
\begin{eqnarray}   \label{eq14}
I_A^{(S)}(q)&=&V_A^2\exp\left(-\frac{q^2 R_A^2}{3}\right)\,,\,\,\,
A\in\{H, D\}\,,
\end{eqnarray}
with the dendrimer volumes \mbox{$V_H=9.9 \pm 0.3$ nm$^3$}, 
\mbox{$V_D=9.7 \pm 0.6$ nm$^3$}, and the radii $R_H=R_D=1.5 \pm 0.2$ nm. 
\cite{rose:02a,rose:02b} Hence the shape function $T_A(r)$ is also a Gaussian 
function implying that these flexible dendrimers have a dense-core structure. 
\cite{ball:04} For the G4-H dendrimers the terms $I_H^{(I)}(q)$ 
and $I_H^{(SI)}(q)$ are small as expected for a rather 
homogeneous scattering length distribution within in the G4-H dendrimers. 
On the other hand, there is a pronounced difference between 
the scattering power of the deuterated end groups and the remaining 
hydrogen atoms in the case of the G4-D dendrimers. Hence, the terms
$I_D^{(I)}(q)$ and $I_D^{(SI)}(q)$ are non-negligible due to the inhomogeneous
scattering length distribution within in the G4-D dendrimers. Figure \ref{fig2}
shows the partial scattering functions $I_H^{(S)}(q) =  I_D^{(S)}(q)$ (dotted line),
$I_H^{(I)}(q)$ (upper solid line), $I_D^{(I)}(q)$ (upper dashed line),
$I_H^{(SI)}(q)$ (lower solid line), and $I_D^{(SI)}(q)$ (lower dashed line)
as obtained from modeling the experimental data. \cite{rose:02a,ding:03} An analysis 
of the Fourier-transform of $I_D^{(I)}(q)$ allows one to calculate the spatial 
distribution function of the deuterated end groups of the G4-D dendrimers. Such an 
analysis has revealed that the end groups are dispersed throughout the dendritic 
molecule. \cite{ball:04,rose:02a}

\subsection{Scattering intensity of interacting dendrimers}
Using the scattering intensities of single dendrimers $I^{(0)}_H(q)$ for G4-H and 
$I^{(0)}_D(q)$ for G4-D [Eq.~(\ref{eq13})] as well as the Ornstein-Zernike 
equations [Eq.~(\ref{eq12})], one can calculate the coherent contribution to the 
scattering intensity $I_{coh}(q)$ for various particle number densities 
according to Eq.~(\ref{eq9}). To this end we have solved the Ornstein-Zernike 
equations together with the hypernetted chain closure

\begin{figure}[t]
\includegraphics[width=7.8cm,clip]{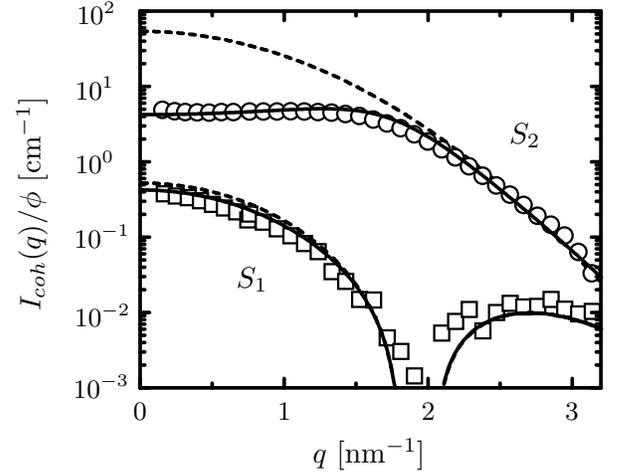}
\caption{Experimentally determined scattering intensities $I_{coh}(q)/\phi$
of binary mixtures of G4-H dendrimers denoted as dendrimers of species $H$ 
[Fig.~\ref{fig1} (A)] and G4-D denrimers denoted as dendrimers of species $D$ 
[Fig.~\ref{fig1} (B)] as function of the magnitude of the scattering 
vector $q$. For sample $S_1$ (open squares) the partial volume fractions 
are $\phi_H=0.12$ and $\phi_D=0.03$, while  $\phi_H=0.15$ and $\phi_D=0$ 
for sample $S_2$ (open circles). The scattering intensities have been normalized 
to the total volume fraction $\phi=\phi_H+\phi_D$ of the dendrimers in the solution. 
Sample $S_1$ has been investigated at the match point of the G4-H dendrimers. 
The solid lines represent the scattering intensity calculated for a solution of 
interacting dendrimers as obtained from  Eq.~(\ref{eq17}) for sample $S_2$ and 
Eq.~(\ref{eq18}) for sample $S_1$. For comparison the dashed lines depict the 
modeling by Eqs.~(\ref{eq17}) and (\ref{eq18}) with $h_{cm}(q)=0$.}
\label{fig3}
\end{figure}
\begin{eqnarray} 
\frac{U_{HD}^{(cm)}(r)}{k_DT}&=&
h_{HD}^{(cm)}(r)-\ln\left(h_{HD}^{(cm)}(r)-1\right)-c_{HD}^{(cm)}(r)
\nonumber
\\&&  \label{eq15}
\end{eqnarray}
and the Gaussian $cm-cm$ interaction potential 

\begin{eqnarray} \label{eq16}
\frac{U_{HD}^{(cm)}(r)}{k_DT}&=& 
N_{HD} \exp\left(-\frac{3 r^2}{4 R^2_{HD}}\right)\,,
\end{eqnarray}
where $N_{HD}$ is a model parameter determining the strength of the 
interaction potential. \cite{liko:02} A Flory-type theory has been used to 
derive this interaction potential on the basis of the experimentally 
determined Gaussian functions $I_H^{(S)}(q)$ and $I_D^{(S)}(q)$ given
by Eq.~(\ref{eq14}). \cite{liko:01,liko:02} Moreover, the hypernetted
chain closure has been found to be in very good agreement with computer
simulation data in the case of soft potentials. \cite{liko:07,lang:00}

Figure \ref{fig3} displays examples of measured and calculated scattering 
intensities of a mixture with the total volume fraction 
$\phi=\phi_H+\phi_D=0.15$ (see Table \ref{table:1}). For the sample $S_1$ (open squares) 
the composition is $\phi_H=\rho_H V_H=0.12$ and $\phi_D=\rho_D V_D=0.03$. In the 
case of the sample $S_2$ (open circles) it is $\phi_H=0.15$ and $\phi_D=0$.
Sample $S_1$ has been investigated at the match point of the G4-H dendrimer, 
i.e., in a solvent where the contrast ${\tilde b}_H$ is zero while 
\mbox{${\tilde b}_D=1.65 \times 10^{10}$ cm$^{-2}$}. This contrast variation 
has been achieved by using mixtures of deuterated and protonated DMA 
as discussed in Ref.~\cite{rose:02a}. For sample $S_2$ (only G4-H) the contrast is 
given by \mbox{${\tilde b}_H=-5.02 \times 10^{10}$ cm$^{-2}$}.  The results of 
the scattering intensities as obtained from Eqs.~(\ref{eq9}) - (\ref{eq16}) with  
\mbox{$V_H=V_D=9.9$ nm$^3$}, $R_{HH}=R_{DD}=R_{HD}=1.4$ nm, and $N_{HH}=N_{DD}=N_{HD}$
(solid lines) are in agreement with the experimental data. For comparison, the 
dashed lines in Fig.~\ref{fig3} depict the modeling of the experimental 
data assuming a mixture of noninteracting dendrimers characterized 
by $h^{(cm)}_{HH}(q)=h^{(cm)}_{DD}(q)=h^{(cm)}_{HD}(q)=0$
in Eq.~(\ref{eq9}). For the sample $S_2$ (only G4-H) the steric interaction between 
the dendrimers that is embodied in the total correlation function 
$h^{(cm)}_{HH}(q)$ leads to a strong depression of the measured 
(open circles) and calculated scattering intensity (upper solid line)
\begin{eqnarray} \label{eq17}
I_{coh}(q)&=&\frac{\phi_H {\tilde b}^2_H}{V_H} I^{(S)}_H(q) 
\left(1+\frac{\phi_H}{V_H} h_{cm}(q)\right)
\end{eqnarray}
as compared to the results for noninteracting dendrimers (upper dashed line).
Here were have defined the \mbox{$cm-cm$} total correlation function 
$h_{cm}(q)\equiv h^{(cm)}_{HH}(q)=h^{(cm)}_{DD}(q)=h^{(cm)}_{HD}(q)$ 
because  $R_{HH}=R_{DD}=R_{HD}$ and $N_{HH}=N_{DD}=N_{HD}$ in Eq.~(\ref{eq16}).
Moreover, we have taken into account that the terms $I_H^{(I)}(q)$ 
and $I_H^{(SI)}(q)$ are very small for the G4-H dendrimers 
as already mentioned. The difference between the solid and dashed line 
in Fig~\ref{fig3} is considerably less pronounced in the case of the scattering 
intensity of the sample $S_1$ although the  $cm-cm$ total correlation function 
$h_{cm}(q)=h_{cm}(q,\phi)$ is the same for both samples. For the sample $S_1$ 
the intensity is given by 
\begin{eqnarray} \label{eq18}
I_{coh}(q)&=&\frac{\phi_D}{V_D}
\left({\tilde b}^2_D I_D^{(S)}(q)+2{\tilde b}_D I_D^{(SI)}(q)+I_D^{(I)}(q)\right)
\nonumber
\\&&\times\left(1+\frac{\phi_D}{V_D} h_{cm}(q)\right)\,.
\end{eqnarray}
The contribution of the  $cm-cm$ total correlation function $h_{cm}(q)$ is 
more pronounced for the sample $S_2$ because the prefactor $\phi_H=0.15$ in the 
last term in Eq.~(\ref{eq17}) is larger than the corresponding prefactor 
$\phi_D=0.03$ in the last term in Eq.~(\ref{eq18}). 

Hence, sample $S_2$ allows one to study 
the mutual interaction between the dendrimers and the validity of the $cm-cm$ 
Ornstein-Zernike equation approach, while the sample $S_1$ provides nearly direct 
information about the shape of the dendrimers at high volume fraction. The fact that 
the calculated scattering intensity (lower solid line) only slightly deviates from 
the experimental data (open squares) indicates that the shape of the dendrimers 
is practically independent of the volume fraction $\phi \leq 0.15$ in contrast 
to pronounced conformational changes of flexible polymers, \cite{daou:75,yeth:91}
bottle-brush polymers, \cite{boli:07,boli:08} and semiflexible polyelectrolytes
\cite{stev:93,yeth:97} in semi-dilute solution. For all systems the contribution 
of the intermolecular interactions to the total free energy increases upon increasing 
the volume fraction. In order reduce this contribution a softening of the stiffness 
of linear macromolecules occurs because for a flexible macromolecule the excluded 
volume that is not available for the other macromolecules is smaller than the 
corresponding one of a rigid macromolecule of the same contour length. As a result 
the size of the macromolecules is reduced considerably upon increasing the volume 
fraction. For example, the radius of gyration of the bottle-brush polymers studied 
in Refs.~\cite{boli:07,boli:08} decreases by the factor $1.8$ upon increasing the 
volume fraction from $0.002$ to $0.04$. In contrast, the flexible and three-dimensional
dendrimers under consideration have to be considered as virtually shape-persistent 
molecules for volume fractions $\phi \leq 0.15$.

\begin{figure}[t]
\includegraphics[width=7.8cm,clip]{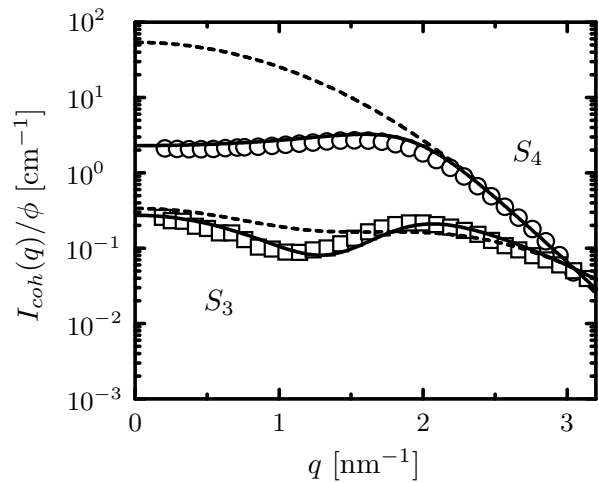}
\caption{Experimentally determined scattering intensities $I_{coh}(q)/\phi$
of binary dendrimer mixtures with $\phi_H=0.03$, $\phi_D=0.2$ for sample $S_3$
(open squares) and $\phi_H=0.23$, $\phi_D=0$ for sample $S_4$ (open circles).
The scattering intensities have been normalized to the total volume fraction
$\phi=\phi_H+\phi_D$. Sample $S_3$ has been investigated at the match point of
the G4-D dendrimers. The solid lines represent the scattering intensity
calculated for a dispersion of interacting dendrimers as obtained from
Eq.~(\ref{eq17}) for sample $S_4$ and Eq.~(\ref{eq20}) for sample $S_3$, while
the dashed lines depict the modeling  with $h_{cm}(q)=0$.
}
\label{fig4}
\end{figure}

Figure \ref{fig4} displays examples of measured and calculated scattering 
intensities of a mixture with the total volume fraction $\phi=\phi_H+\phi_D=0.23$ 
with $\phi_H=0.03$, $\phi_D=0.2$ in the case of the sample $S_3$ (open squares)
and $\phi_H=0$, $\phi_D=0.23$ in the case of the sample $S_4$ (open circles).
The total volume fraction of the dendrimers in these samples is comparable to 
the overlap volume fraction. \cite{goet:05}
Sample $S_3$ has been investigated at the match point of the G4-D dendrimer, 
i.e., in a solvent where the contrast ${\tilde b}_D$ is zero while 
\mbox{${\tilde b}_H=-1.65 \times 10^{10}$ cm$^{-2}$}. For sample $S_4$ (only G4-H) 
the contrast is given by \mbox{${\tilde b}_H=-5.02 \times 10^{10}$ cm$^{-2}$}. As 
before the nearly agreement of the open circles with the results as obtained from 
the Ornstein-Zernike approach and the decoupling approximation (upper solid line) 
indicates that the Gaussian $cm-cm$ interaction potential [Eq.~(\ref{eq16})] 
is appropriate even at high volume fractions. In the case of sample $S_3$
all terms in Eq.~(\ref{eq9}) contribute to the scattering intensity according to
\begin{eqnarray}
\lefteqn{I_{coh}(q)=}\nonumber
\\&&\frac{\phi_H {\tilde b}^2_H}{V_H} I_H^{(S)}(q)+
\frac{\phi_D}{V_D} I_D^{(I)}(q)\nonumber
\\&&+
\left(\frac{\phi_H}{V_H} \sqrt{{\tilde b}^2 I_H^{(S)}(q)} + 
\frac{\phi_D}{V_D} \sqrt{I_D^{(I)}(q)} \right)^2 h_{cm}(q)\,.\label{eq20}
\end{eqnarray}
The shape of $I_{coh}(q)$ is mainly determined by the functions $I_D^{(I)}(q)$ 
and $h_{cm}(q)$ characterizing the distribution of the deuterated end groups of 
the G4-D dendrimers and the intermolecular correlations, respectively, because 
$\phi_D \gg \phi_H$. The small deviations between the lower solid line and the 
open squares in Fig.~\ref{fig4} might be either due to slight changes of the 
distribution of the end groups as compared to the noninteracting system in 
dilute solution or due to the fact that the calculated $cm-cm$ total 
correlation function is not in exact agreement with the experimental data as 
is apparent from sample $S_4$. However, possible deformations of the 
dendrimers are very weak as compared to the abovementioned linear macromolecules. 
For comparison we note that a computer simulation study has also revealed only 
very small deformations of flexible dendrimers, i.e., the overall size of 
those dendrimers decreases by less than 2 \% upon increasing the volume fraction 
from infinite dilution to the overlap volume fraction. \cite{goet:05} 
Thus, we conclude that even in the range of the overlap concentration the
dendrimers preserve their structure.

Finally the results shown in Fig.~\ref{fig3} and \ref{fig4} demonstrate that the
decoupling approximation [Eq.~(\ref{eq9})] may be used as a first approximation 
although the height of the calculated peak of the scattering intensity at 
$q \approx 1.85$ nm for sample $S_4$ is slightly higher than the experimental data. 
A similar overestimation of the height of such peaks has already been observed 
by comparing the results of the decoupling approximation with exact evaluations of 
the scattering intensity as obtained from a computer simulation study of monodisperse 
dendrimers. \cite{goet:05}

\vspace*{0.5cm}
\section{Summary}
A general analysis of the scattering intensities of a binary mixture of dissolved 
dendrimers obtained by small-angle neutron scattering has been presented. Partially 
deuterated dendrimers [Fig.~\ref{fig1} (B)] have been mixed with the same but protonated 
dendrimers [Fig.~\ref{fig1} (A)] in dimethylacetamide which is a good solvent for this 
system. Contrast variation has been used in order to obtain partial scattering intensities
allowing one to investigate both the structure and the interaction of dendrimers in 
concentrated solution. An interpretation of the partial scattering intensities leads 
to the conclusion that the overall structure of both dendrimers is identical and an 
appreciable number of end groups must fold back into the interior of the dendrimers
in dilute solution [Fig.~\ref{fig2}]. 

Binary dendrimer mixtures have been investigated at various contrasts [Table \ref{table:1}] 
in order to distinguish the influence of the shape of individual dendrimers from the 
influence of intermolecular interactions on the scattering intensities at high concentrations.
The samples $S_1$ and $S_3$ provide nearly direct information about the shape of the 
dendrimers at high concentration. An analysis of the measured scattering intensities for 
both samples [Figs.~\ref{fig3} and \ref{fig4}] has revealed that the shape of the dendrimers 
is virtually independent of the concentration. Hence, the flexible poly(propyleneamine) 
dendrimers of fourth generation have to be considered as shape-persistent molecules within the present limits of approximation.
The samples $S_2$ and $S_4$ allows one to study the mutual interaction between the 
dendrimers at two different concentrations. A comparison of the measured 
scattering intensities with the calculated ones [Figs.~\ref{fig3} and \ref{fig4}] 
demonstrate that the center-of-mass Ornstein-Zernike relations [Eq.~(\ref{eq12})]
together with the hypernetted chain closure [Eq.~(\ref{eq15})] and the Gaussian
interaction potential [Eq.~(\ref{eq16})] may be used as a good approximation although
the height of the calculated peak of the scattering intensities are slightly higher than
the experimental data.

\section{Acknowledgement}
The authors thank the Institute Laue-Langevin in Grenoble (France) for providing
beamtime at the Instrument D11.

\end{document}